\documentclass[12pt, a4paper]{iopart}

\usepackage[]{graphicx}  
\usepackage{wrapfig}
\usepackage{courier}
\usepackage{multirow}
\usepackage{cite}
\usepackage[rightcaption]{sidecap}

\usepackage{xcolor}

\begin{document}

\title[]{Radial electric field and density fluctuations measured by Doppler reflectometry during the post-pellet enhanced confinement phase in W7-X }

\author[1]{T. Estrada$^1$, D. Carralero$^1$, T. Windisch$^2$, E. S\'anchez$^1$, J. M. Garc\'ia-Rega\~na$^1$, J. Mart\'inez-Fern\'andez$^1$, A. de la Pe\~na$^1$, J. L. Velasco$^1$, J. A. Alonso$^1$, M. Beurskens$^2$, S. Bozhenkov$^2$, H. Damm$^2$, G. Fuchert$^2$, R. Kleiber$^2$, N. Pablant$^3$, E. Pasch$^2$, and the W7-X team$^2$}

\address{$^1$ Laboratorio Nacional de Fusi\'on. CIEMAT, 28040 Madrid, Spain}
\address{$^2$ Max-Plank-Institut f\"ur Plasmaphysik, D-17491 Greifswald, Germany}
\address{$^3$ Princeton Plasma Physics Laboratory, Princeton, NJ, USA}

\ead{teresa.estrada@ciemat.es}


\begin{abstract}
{

Radial profiles of density fluctuations and radial electric field, $E_r$, have been measured using Doppler reflectometry during the post-pellet enhanced confinement phase achieved, under different heating power levels and magnetic configurations, along the 2018 W7-X experimental campaign. A pronounced  $E_r$-well is measured with local values as high as -40 kV/m in the radial range $\rho \sim 0.7-0.8$ during the post-pellet enhanced confinement phase. 
The maximum  $E_r$ intensity scales with both plasma density and Electron Cyclotron Heating (ECH) power level following a similar trend as the plasma energy content. 
A good agreement is found when the experimental $E_r$ profiles are compared to simulations carried out using the neoclassical codes DKES and KNOSOS. 
The density fluctuation level decreases from the plasma edge toward the plasma core and the drop is more pronounced in the post-pellet enhanced confinement phase than in reference gas fuelled plasmas. Besides, in the post-pellet phase, the density fluctuation level is lower in the high iota magnetic configuration than in the standard one. In order to discriminate whether this difference is related to the differences in the plasma profiles or in the stability properties of the two configurations, gyrokinetic simulations have been carried out using the codes \texttt{stella} and EUTERPE. The simulation results point to the plasma profile evolution after the pellet injection and the stabilization effect of the radial electric field profile as the dominant players in the stabilization of the plasma turbulence.    

}
\end{abstract}

\section{Introduction}
 
High core plasma densities will be essential in future fusion reactors in order to maximize fusion power. In present day experiments, both tokamaks and helical devices, high central densities are achieved using intense high repetition rate pellet injection, extending the operational regime beyond the density limit predicted by the empirical scalings~\cite{Ohyabu:2006,Sakamoto:2009,Lang:2012,Lang:2018}. 
Improved plasma performance has been observed associated to core plasma fuelling by pellet injection in many experiments during the last decades~\cite{Greenwald:1984,Jacquinot:1988,Kamada:1989,Smeulders:1995,Milora:1995,Ida:1996,Maget:1999,Sakamoto:2001,Gohil:2003,Romanelli:2004,Pegourie:2007,Takeiri:2017,McCarthy:2017,Motojima:2019}. In most experiments, a reduction of the ion transport and an increase in the core ion temperature are found associated to peaked density profiles. The energy confinement is observed to increase over similar discharges fuelled only by gas puffing. 
Similar behaviour has been reported associated to deep plasma fuelling by Neutral Beam Injection (NBI)~\cite{Stroth:1998a,Ida:1999,Baldzuhn:2000}.
Accompanying the increase in the plasma density and temperature gradients,  a strong negative radial electric field, $E_r$, has been measured in a radial region that extends deep into the plasma bulk~\cite{Ida:1996,Maget:1999,Gohil:2003,Stroth:1998a,Ida:1999}. 
The reduction in the ion transport has been interpreted to be due to the stabilization of the ion temperature gradient (ITG) driven instabilities by the density peaking~\cite{Romanelli:2004}. Besides, the impact of the $E_r$-shear on plasma turbulence has been shown to play a role too on the enhanced plasma performance~\cite{Maget:1999}.

In the stellarator W7-X, a notable improvement in the plasma performance has been observed after the injection of a series of frozen hydrogen pellets into ECH heated plasmas~\cite{Klinger:2019,Baldzuhn:2020,Bozhenkov:2020}. In these experiments, the pellet series rise the plasma density considerably and the central fuelling results in a peaked density profile. At the end of the pellet series, the ion and electron temperatures rise and almost equilibrate, and the plasma stored energy increases by more than 40\% reaching values above 1 MJ. 
The $E_r$ profile measured by X-ray Imaging Crystal Spectrometer (XICS) diagnostic indicates that ion-root conditions are achieved in the whole plasma column~\cite{Pablant:2020}. 
Power balance analysis indicates that during the enhanced  confinement phase the turbulent transport is reduced getting close to the neoclassical level~\cite{Bozhenkov:2020}.   
This enhanced confinement phase lasts for several confinement times and terminates as the density and its peaking decay. The stabilisation of this enhanced confinement will be of the utmost importance for exploring reactor-relevant scenarios in the W7-X next operational phase with an actively cooled divertor. 
A candidate to explain these observations is the reduction of the turbulent transport due to both, the stabilization by centrally peaked density profile of the ion temperature gradient (ITG) driven instabilities, and the specific stability properties of the electron-density-gradient driven trapped electron mode (TEM) in the W7-X magnetic geometry~\cite{Proll:2012,Alcuson:2020,Xanthopoulos:2020}. This theoretical description has been to some extent supported by the drop in the line-integrated density fluctuation level measured with a phase contrast imaging (PCI) system during the enhanced confinement phase~\cite{Stechow:2020}. 

The present work reports, for the first time, radially-resolved measurements of density fluctuations and radial electric field by Doppler reflectometry (DR), that can provide new insights into the nature of this transient suppression of turbulence. The study encompasses several high-performance cases, observed under different heating power levels and magnetic configurations along the 2018 W7-X experimental campaign. A comparison with both, neoclassical and gyrokinetic simulations, is also presented in an attempt to test the ability of our most sophisticated tools to reproduce these observations and eventually disentangle the impact of $E_r$, plasma kinetic profiles, and magnetic configuration on the density fluctuations. 

Measurements are obtained with a DR system working in the 50-75 GHz frequency range in O-mode polarization and with fixed probing beam angle of $\alpha = 18^\circ$~\cite{Windisch:2019,Carralero:2020}. Under these conditions, perpendicular wave-numbers of the turbulence in the range $k_\perp \sim 7-10$ cm$^{-1}$ are measured at the accessible local densities in the range from 2.8 to 6.3 $\times 10^{19}$ m$^{-3}$. 

The remainder of the paper is organised as follows. The experimental set-up is described in section 2 and the experimental results in section 3: those related to the radial electric field measurements in section 3.1 including the comparison with the neoclassical predictions in subsection 3.1.1, and those describing the density fluctuations in section 3.2, and the comparison with gyrokinetic simulations in subsection 3.2.1. 
Finally, the summary is included in section 4.

\section{Experimental set-up}

W7-X is an optimised superconducting stellarator with major radius $R = 5.5$ m, minor radius $a = 0.5$ m and magnetic field $B_0 = 2.5$ T~\cite{Klinger:2016,Wolf:2017}.
W7-X, with 50 non-planar and 20 planar superconducting coils, offers the possibility to explore different magnetic configurations~\cite{Geiger:2014}. Besides, island divertor operation is possible with edge rotational transform equal to 5/6 (low iota), 5/5 (standard) and 5/4 (high iota configuration)~\cite{Klinger:2019}. 

In the present experiments, plasmas are created and heated by ECH $2^{nd}$ harmonic at 140 GHz. The ECH system~\cite{Erckmann:2007} consists of 10 long-pulse gyrotrons with a power per gyrotron of up to 0.8 MW, coupled into the plasma either in X-mode or in O-mode polarization. The later allows expanding the operational plasma density range beyond the X-mode cut-off density ($1.2 \times10^{20}$ m$^{-3}$).

The pellet injection system in operation at W7-X during the 2017-2018 experimental campaigns can provide series of hydrogen ice pellets with a variable repetition frequency~\cite{Baldzuhn:2019}. 
During these campaigns, long series of pellets with more than 30 pellets per discharge/program were performed reaching densities of up to 1.4 $\times$ 10$^{20}$ m$^{-3}$.
These experiments demonstrate central fuelling during pellets series and a better fuelling efficiency and a deeper penetration with the number of pellet within a series. Initial pellets in the series cool the plasma edge and facilitate a deeper penetration for later pellets. 

The main experimental results presented in this work have been obtained using Doppler reflectometry (DR). This technique is used to measure the density fluctuations and its perpendicular rotation velocity, at different spatial scales and with good spatial and temporal resolution~\cite{Hirsch:2001,Hennequin:2004,Conway:2004,Happel:2009}.
In these experiments a V-band DR system working in O-mode polarization has been used to measure radially-resolved density fluctuations and $E_r$ profiles~\cite{Carralero:2020}. The reflectometer front end, installed at port AEA21 (toroidal angle $\phi = 72 ^\circ$), uses a single antenna and a set of mirrors for launching and receiving the signal at fixed probing beam angle of $\alpha = 18^ \circ$~\cite{Windisch:2019}. The fixed probing beam angle limits the accesible perpendicular wave-numbers of the turbulence within the range $k_\perp \sim 7-10$ cm$^{-1}$. 
During each experimental program, the probing frequency of the reflectometer is scanned in a hopping mode from 50 to 75 GHz, typically in steps of 1 GHz, 10 ms long. Thus, every 250 ms a complete frequency scan is performed probing local densities from 2.8 to 6.3 $\times 10^{19}$ m$^{-3}$. The corresponding radial positions and perpendicular wave-numbers are calculated using the 3D ray-tracing code TRAVIS~\cite{Marushchenko:2014} with the proper magnetic configuration VMEC equilibrium and the density profile measured by the Thomson Scattering diagnostic~\cite{Pasch:2016}. 
Uncertainties are calculated as the dispersion of the radial positions and perpendicular wave-numbers calculated using a bundle of rays that simulates the DR probing beam.

The perpendicular rotation velocity of the plasma turbulence measured by DR is a composition of both the plasma $E\times B$ velocity and the intrinsic phase velocity of the density fluctuations: $u_\perp = v_{E\times B} + v_{ph}$. In cases in which the condition $v_{E\times B} \gg v_{ph}$ holds, $E_r$ can be obtained directly from the perpendicular rotation velocity: $E_r = u_\perp \cdot B =  2 \pi f_D/k_\perp  \cdot B$, where $f_D$ is the frequency of the Doppler peak and B is the magnetic field. 
Experiments performed both in tokamaks and stellarators have demonstrated that this assumption is in general valid~\cite{Hirsch:2001,Conway:2004,Estrada:2009,Estrada:2019}. In high collisionality plasmas, however, the contribution of $v_{ph}$ to $u_\perp$ may become relevant~\cite{Vermare:2011,Estrada:2019}.
In the present experiments, it is assumed that $v_{E\times B} \gg v_{ph}$ in order to calculate $E_r$ and, for a number of cases, the obtained $E_r$ profile is compared with that obtained using the neoclassical codes DKES (Drift Kinetic Equation Solver)~\cite{Hirshman:1986} and KNOSOS (KiNetic Orbit-averaging SOlver for Stellarators)~\cite{Velasco:2020}. A good agreement between them would support the assumption that the perpendicular rotation velocity of the plasma turbulence is dominated by the $E\times B$ velocity.

The experimental $E_r$ values measured at different flux surfaces correspond to the local values measured by DR at the outboard midplane of the bean-shaped plane. In this sense, the term local $E_r$ is used to refer to experimental $E_r$. As it is explained below, the flux surface expansion has to be taken into account when these local values are compared with those calculated by neoclassical codes.

Regarding the density fluctuations, the power of the back-scattered DR signal is the relevant quantity proportional to $| \delta n|^2$ and given by $S=A_D \cdot \Delta f_D$, where $A_D$ and $\Delta f_D$ are the height and width of the Doppler peak.
It has to be noted that, in general, a microwave generator working with variable frequency produces a different power output at each frequency. Besides, the transmitted power through the transmission line may also depend on the frequency. Therefore, a power calibration of the Doppler reflectometer is indispensable for a proper comparison of the fluctuations measured at different frequencies. The calibration applied to the data presented in section 3.2 is a combination of a direct calibration of the ex-vessel transmission line components in fundamental waveguide, the theoretical behaviour of the in-vessel directional coupler and the result of a systematic error analysis.  The last term yields rather small values ($< 2$ dB for most probing frequencies), however, it is worth to consider it as it accounts for the transmitted power frequency dependence of the components not included in the calibration. 
Finally, it has to be noted that the configuration of the Doppler reflectometer hardware was fixed during the whole experimental campaign what allows using the same power calibration factors. Thus, fluctuation levels measured at different scenarios can be directly compared.

\section{Experimental results}

The enhanced confinement phase observed in W7-X after the injection of pellets series has been described in recent publications~\cite{Klinger:2019,Stechow:2020,Baldzuhn:2020,Bozhenkov:2020}. Core plasma fuelling by the injection of pellets allows rising the central density, increasing considerably the density peaking. After the pellet phase, the density and its peaking gradually decay while the electron and ion temperature increase and almost equilibrate, giving rise to a pronounced increase in the diamagnetic energy.  
This high performance phase lasts for several confinement times and terminates as the density and its peaking finally decay. 
As discussed in~\cite{Bozhenkov:2020}, this confinement improvement cannot be explained only by the increase in the plasma density: gas fuelled plasmas with similar line-averaged density and heating power have lower ion temperature and plasma energy. 
Although transient, the enhanced confinement phase lasts, in most cases, long enough to perform a full DR frequency scan, essential to measured the radial profiles of $E_r$ and density fluctuations. These profiles are compared with those measured in gas fuelled plasmas with similar density and heating power level. 
This study includes several high performance phases, observed under different heating power levels and magnetic configurations, along the 2018 W7-X experimental campaign.

\subsection{Radial electric field} \label{Radial electric field}

\begin{figure}[t]
 \center
 \includegraphics[width=0.50\columnwidth,trim= 0 0 0 0]{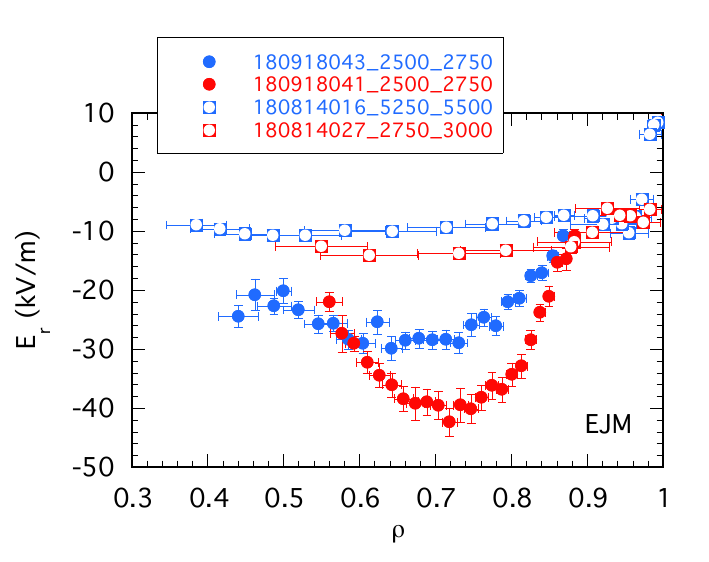}     
  \includegraphics[width=0.23\columnwidth,trim= 0 -60 0 0]{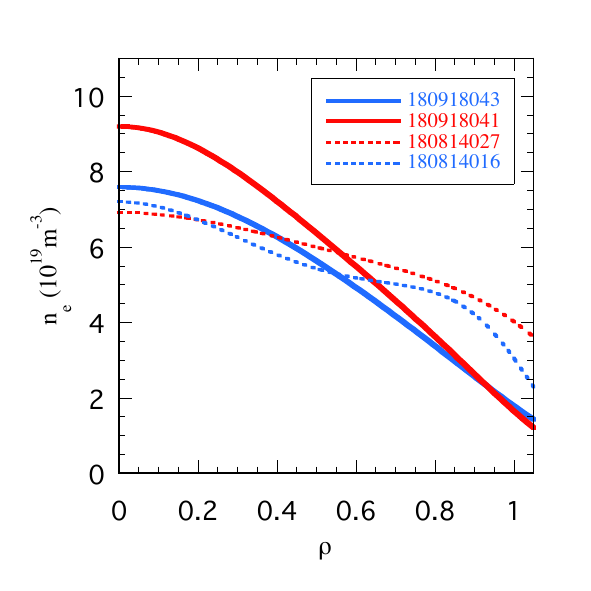}
 \includegraphics[width=0.23\columnwidth,trim= 0 -60 0 0]{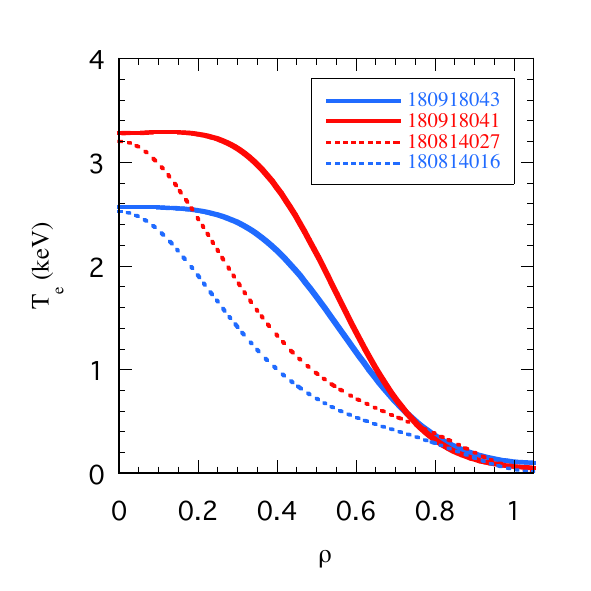}
 \caption{Left: Radial profiles of $E_r$ measured during the post-pellet phase in two experimental programs (solid symbols) and those measured in gas fuelled reference plasmas (open symbols), with $n_e \sim 8 \times 10^{19}$ m$^{-2}$ and $P_{ECH} =$ 3 MW (in blue), and $n_e \sim 9 \times 10^{19}$ m$^{-2}$ and $P_{ECH} =$ 5.5 MW (in red). Right: Corresponding electron density and temperature profiles measured by Thomson Scattering during the post-pellet phase in the two pellet fuelled plasmas (solid lines) and in the reference ones (dotted lines).} 
\label{f:fig_1}
\end{figure}

Radial electric field profiles have been measured using DR in experimental programs with different pellet sequence reaching different plasma densities and heated with different ECH power levels, in two magnetic configurations: standard and high iota, named EJM and FTM, respectively. Details on these configurations can be found in~\cite{Klinger:2019,Carralero:2020}. 
A pronounced $E_r$-well is measured with local $E_r$ values as high as $-40$ kV/m in the radial range $\rho \sim 0.7-0.8$ during the post-pellet enhanced confinement phase in both magnetic configurations.  In plasmas with similar ECH heating power and density but fuelled by gas puffing, the local $E_r$ profile is rather flat with values close to $-15$ kV/m within the range $\rho \sim 0.5-0.9$. 
As an example, figure  \ref{f:fig_1} shows the local $E_r$ profiles measured during the post-pellet enhanced confinement phase in two cases with different plasma density and ECH heating power in the standard magnetic configuration. 
For comparison, the $E_r$ profiles measured in plasmas with similar density and ECH power but fuelled by gas puffing are also shown in figure \ref{f:fig_1}. 
The corresponding electron density and temperature profiles are shown in figure  \ref{f:fig_1}.right. 
The vertical error bars in figure \ref{f:fig_1}.left reflect both the uncertainty on $k_\perp$ and the error on the fit of the Doppler peak, while the horizontal error bars reflects the uncertainty on the radial position calculated with TRAVIS but do not include uncertainties on the density profiles. The later may impact the precise radial localization of the DR measurements specially at the plasma edge but do not modify the findings and conclusions reported in this work. 

\begin{figure}[h]
\center
\includegraphics[width=0.32\columnwidth,trim= 20 0 0 0]{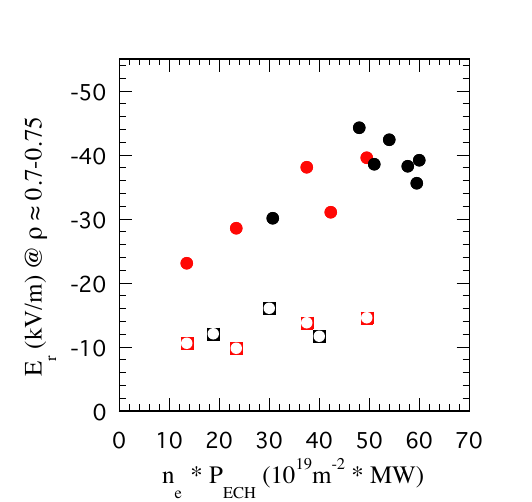}
\includegraphics[width=0.32\columnwidth,trim= 20 0 0 0]{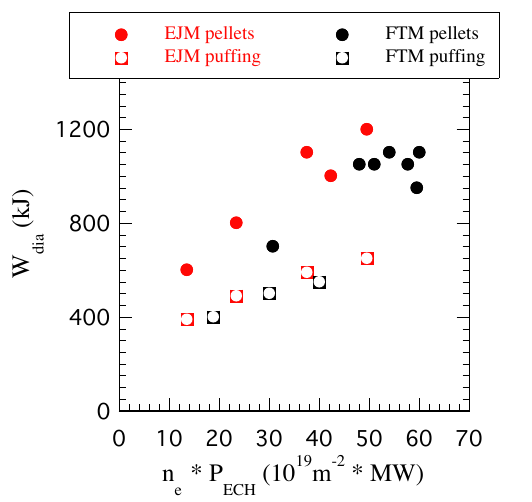}
\includegraphics[width=0.32\columnwidth,trim= 20 0 0 0]{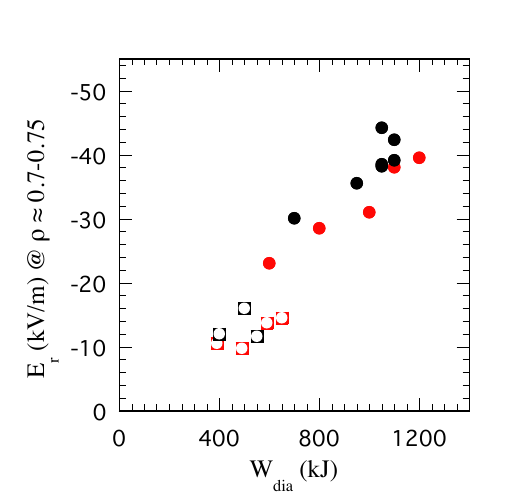}
\caption{Local $E_r$ intensity measured by DR at $\rho \sim$ 0.7-0.75 (left) and diamagnetic energy (centre) during the post-pellet high confinement phase (solid symbols) and in gas fuelled plasmas (open symbols) as a function of the product $n_e \times P_{ECH}$.  The relation between the $E_r$ intensity and the diamagnetic energy is shown in the right panel. Standard magnetic configuration EJM is represented in red and high iota configuration FTM in black.}
\label{f:fig_2}
\end{figure}

The plasma with higher heating power and density (program$\#$ 180918041) has more pronounced plasma gradients and deeper $E_r$ well in the post-pellet phase
The analysis of a number of scenarios confirm this observation.
The maximum $E_r$ intensity at the $E_r$-well scales with both density and ECH power level following a similar trend as the plasma energy content.   
The local $E_r$ intensity measured by DR at $\rho \sim$ 0.7 - 0.75 during the post-pellet high confinement phase  is shown in figure \ref{f:fig_2}.left (solid symbols) as a function of the product: plasma density $\cdot$ ECH power. The data correspond to twelve discharges, five in the standard magnetic configuration and seven in the high iota configuration. Besides, data from gas fuelled discharges in both configurations is also included as a reference (open symbols). For the same discharges, the diamagnetic energy, $W_d$, is shown in figure  \ref{f:fig_2}.centre and the relation between $E_r$ and $W_d$ in figure  \ref{f:fig_2}.right.
Both, the maximum $E_r$ intensity and $W_d$ are much higher in the post-pellet phase than in the gas fuelled plasmas for similar density and heating power level. 
Besides, as can be seen in figure  \ref{f:fig_2}, there are almost no differences between the two magnetic configurations. For given values of plasma density and heating power, the local $E_r$ intensity at the post-pellet phase is similar in both configurations. As it will shown in the next section, this is not the case when the density fluctuation level is compared.

\subsubsection{Comparison with neoclassical simulations.}

\begin{figure}[t]
\center
\includegraphics[width=0.42\columnwidth,trim= 0 10 0 20]{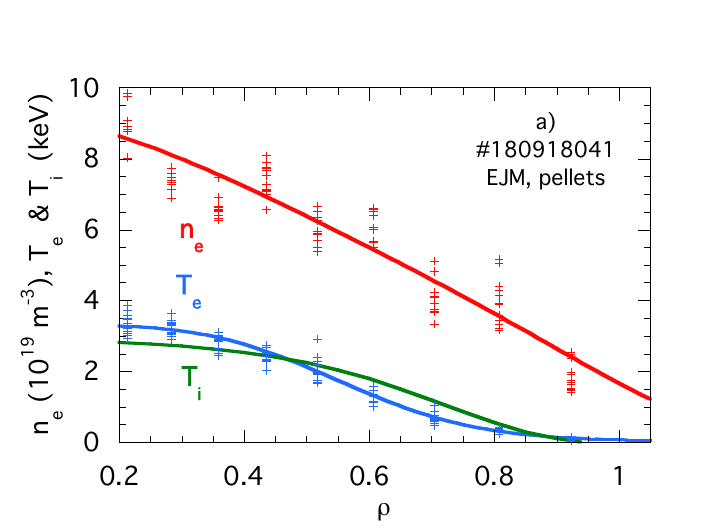}
\includegraphics[width=0.42\columnwidth,trim= 0 10 0 20]{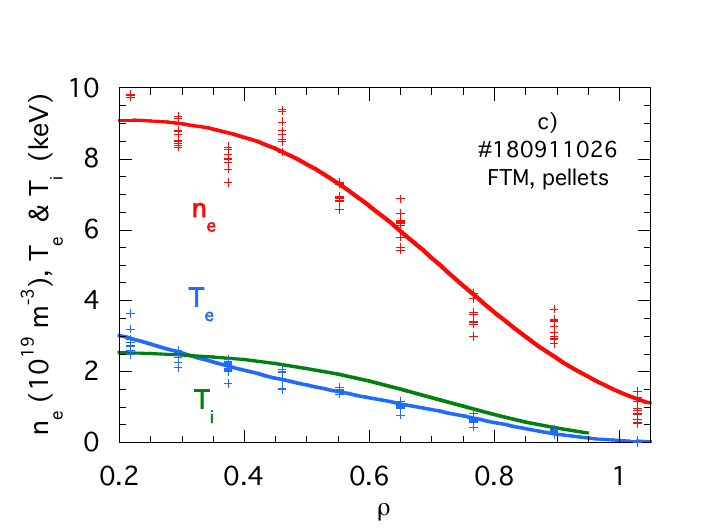}
\includegraphics[width=0.42\columnwidth,trim= 0 10 0 20]{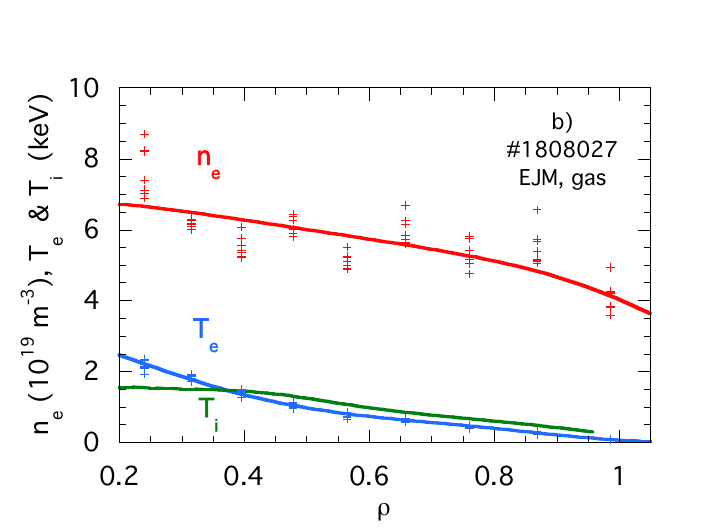}
\includegraphics[width=0.42\columnwidth,trim= 0 10 0 20]{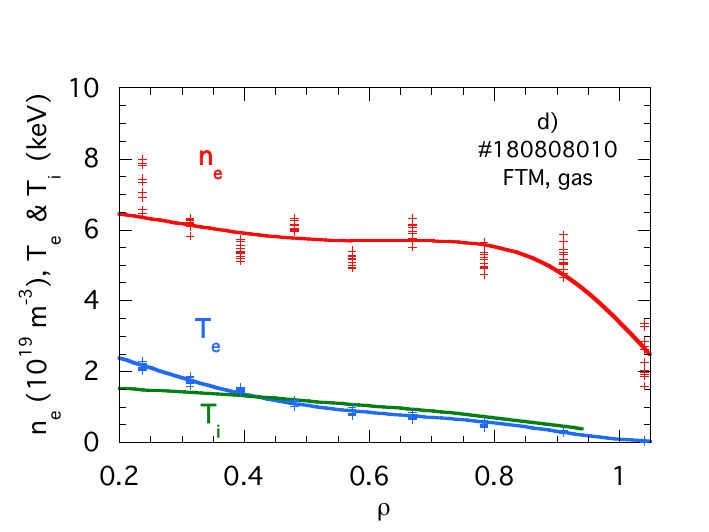}
\caption{Electron density and temperature profiles measured by Thomson Scattering diagnostic and ion temperature profile measured by XICS diagnostic during the post-pellet phase in the standard EJM (a) and high iota FTM magnetic configuration (c), and in the reference gas fuelled plasmas in the two configurations (b) and (d), respectively. The solid lines represent the fits to the experimental values (symbols) measured during the time interval of the DR analysis.}
\label{f:fig_3}
\end{figure}

The $E_r$ profiles measured using DR during the post-pellet enhanced confinement phase have been compared with the neoclassical predictions obtained from the codes DKES and KNOSOS  (the monoenergetic approach is employed; DKES is employed for calculating the transport coefficients at high collisionalities and KNOSOS at low collisionalities, see~\cite{Velasco:2020} for more details). 
Two scenarios have been selected with similar density ($\sim 9 \times 10^{19}$ m$^{-2}$) and heating power ($\sim$ 5.5 - 6.0 MW) the first in the standard EJM magnetic configuration and the second in the high iota FTM configuration. In both cases the local $E_r$ profile measured by DR in the post-pellet phase is very similar with maximum $E_r$ at the $E_r$-well close to -40 kV/m. As before, plasmas with similar density and heating power but fuelled with gas puffing are also studied as a reference. 
For the calculations, the electron density and temperature profiles measured by Thomson Scattering diagnostic~\cite{Pasch:2016} and the ion temperature profile measured by X-ray Imaging Crystal Spectrometer (XICS) diagnostic~\cite{Pablant:2018} are taken as an input.
These profiles are shown in figure  \ref{f:fig_3}. The profiles measured during the post-pellet phase are shown in the top panels and those in the reference gas fuelled plasmas in the bottom panels, in the standard magnetic configuration on the left and in the high iota configuration on the right. Note that for the simulations, $T_i$ is set equal to $T_e$ at the plasma periphery.

\begin{figure}[h]
\center
\includegraphics[width=0.49\columnwidth,trim= 0 0 10 0]{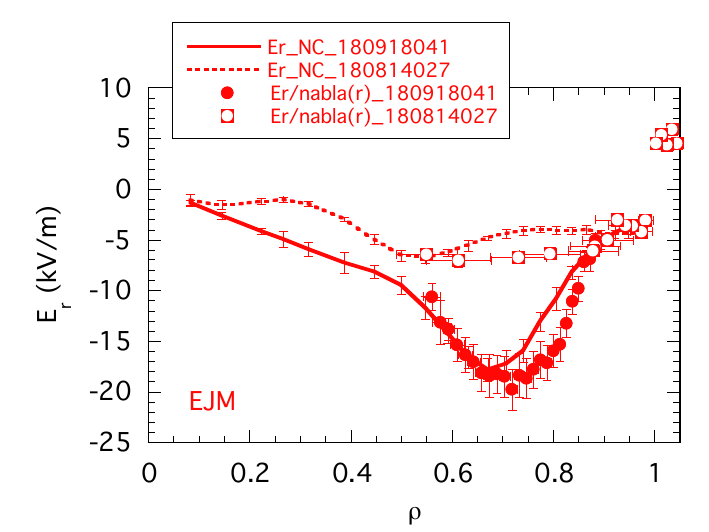}
\includegraphics[width=0.49\columnwidth,trim= 0 0 10 0]{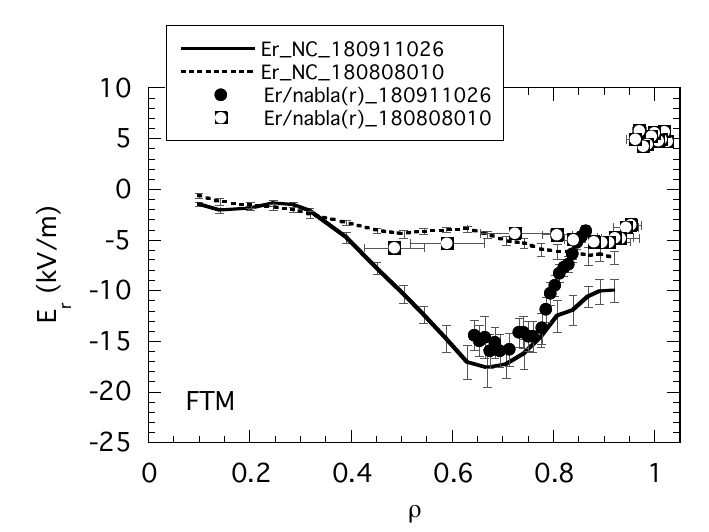}
\caption{Comparison of experimental $E_r$ profiles normalised with $| \nabla r|$ (symbols) and neoclassical predictions obtained using the codes DKES and KNOSOS (lines) for the post-pellet enhanced confinement phase (solid symbols and thick lines) and for the corresponding gas fuelled plasmas (open symbols and broken lines); for the standard magnetic configuration EJM (left, in red) and for the high iota FTM configuration (right, in black). The error bars in the simulated $E_r$ profiles result from a sensitivity study performed assuming deviations of 10$\%$ from the measured density and temperature profiles and their gradients.}
\label{f:fig_4}
\end{figure}

In order to compare the experimental $E_r$ profiles with the simulations, the flux surface expansion has to be taken into account and the experimental local $E_r$ profiles have to be normalised with $| \nabla r|$. At the position of the DR measurements (roughly at the outboard midplane of the bean-shaped plane), the flux surfaces are highly elongated with a rather high flux surface compression: $| \nabla r| \sim 2.1$ in the standard magnetic configuration and $| \nabla r| \sim 2.6$ in the high iota configuration. 

The comparison of experimental $E_r$ profiles and neoclassical predictions is shown in figure \ref{f:fig_4}. 
Taking into account the possible error sources coming from possible uncertainties in the fits of the plasma profiles and from the assumption $v_{E\times B} \gg v_{ph}$ that allows to obtain $E_r$ directly from the perpendicular rotation velocity as $E_r = u_\perp B$, a remarkable good agreement is found in the four cases.
The flat $E_r$ profiles measured in the gas fuelled plasmas, and the $E_r$-well radial position and intensity measured during the post-pellet phase, are rather well reproduced by the neoclassical simulations.

\subsection{Density Fluctuations} \label{Density Fluctuations}

\begin{figure}[h]
\center
\includegraphics[width=0.49\columnwidth,trim= 0 0 20 0]{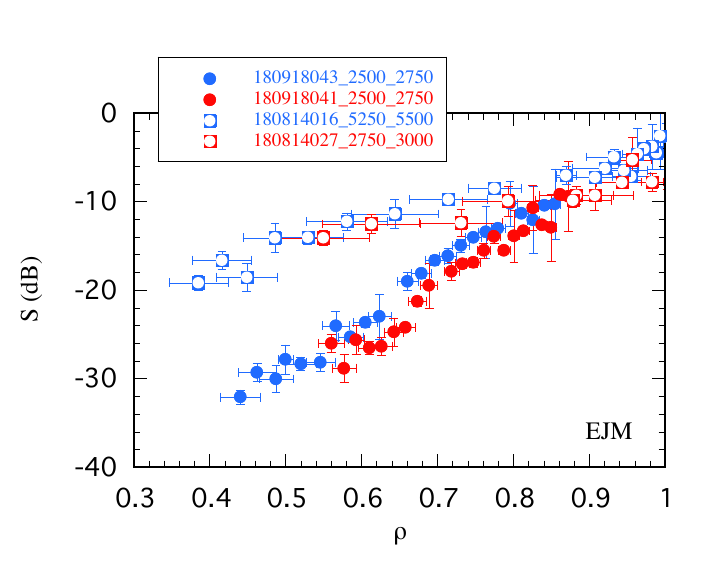}
\caption{Radially resolved density fluctuations measured during the post-pellet phase in two experimental programs (solid symbols) and those measured in gas fuelled reference programs (open symbols), with $n_e \sim 8 \times 10^{19}$ m$^{-2}$ and $P_{ECH} =$ 3 MW (in blue), and $n_e \sim 9 \times 10^{19}$ m$^{-2}$ and $P_{ECH} =$ 5.5 MW (in red). Same experimental programs as in figure \ref{f:fig_1}.}
\label{f:fig_5}
\end{figure}

 Radially resolved density fluctuations are obtained from the power of the back-scattered DR signals measured in the experimental programs presented in the previous section.  
In general, the density fluctuation level decreases from the plasma edge toward the plasma core in all plasma scenarios but the drop is more pronounced in the post-pellet enhanced confinement phase than in the gas fuelled plasmas. 
To illustrate this radial dependence, figure \ref{f:fig_5} shows the density fluctuations measured in the experimental programs presented in figure \ref{f:fig_1}: standard magnetic configuration with $n_e \sim 8 \times 10^{19}$ m$^{-2}$ and $P_{ECH} =$ 3 MW (in blue), and $n_e \sim 9 \times 10^{19}$ m$^{-2}$ and $P_{ECH} =$ 5.5 MW (in red).
In the post-pellet phase (solid symbols) the fluctuation level decreases by about $15 - 20$ dB within the radial range from  $\rho \sim 0.9$ to  $\rho \sim 0.5$, while it decreases only by about 10 dB in the gas fuelled plasmas (open symbols). 
The lower fluctuation level measured during the post-pellet phase may be the result of a change in the turbulence nature or intensity linked to the change in the plasma profiles and/or the result of the $E_r$-shear impact on turbulence. 
Gyrokinetic simulations have been performed aiming to asses the relative importance of the two effects (see subsection 3.2.1).

\begin{figure}[h]
\center
\includegraphics[width=0.49\columnwidth,trim= 0 0 20 0]{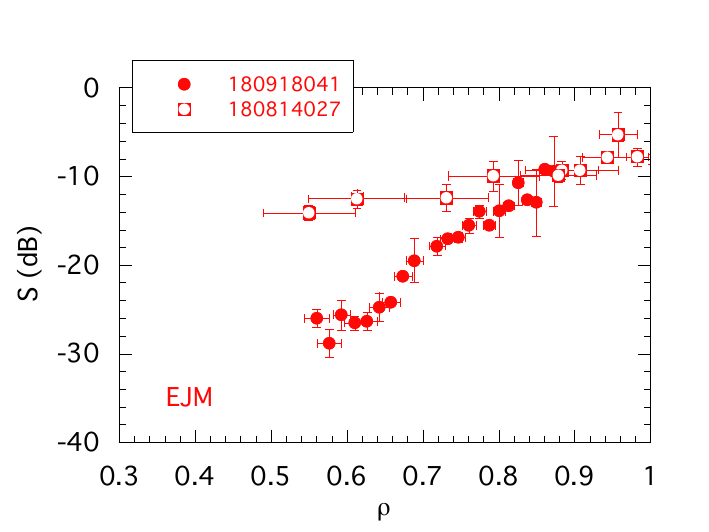}
\includegraphics[width=0.49\columnwidth,trim= 0 0 20 0]{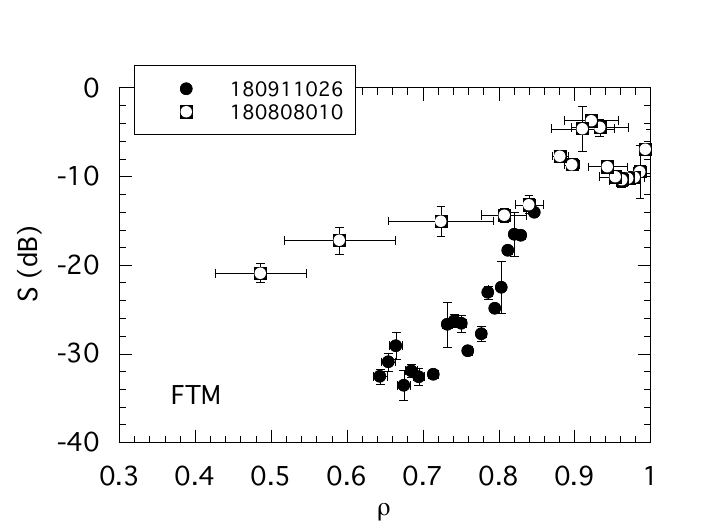}
\caption{Radially resolved density fluctuations measured during the post-pellet phase (solid symbols) and those measured in gas fuelled reference programs (open symbols), in the standard magnetic configuration (left, in red) and in the high iota configuration (right, in black). Same experimental programs as in figures \ref{f:fig_3} and \ref{f:fig_4}.}
\label{f:fig_6}
\end{figure}

Significant differences are found when the two magnetic configurations are compared as regards the density fluctuations. The density fluctuation level during the post-pellet enhanced confinement phase is lower in the high iota magnetic configuration than in the standard one in the radial range $\rho \sim$ 0.6 - 0.8. Figure \ref{f:fig_6} shows the fluctuation level during the post-pellet phase and in gas fulled plasmas measured in the two magnetic configurations: standard EJM (left, in red) and high iota FTM (right, in black). These cases correspond to the four experimental programs represented in figures \ref{f:fig_3} and \ref{f:fig_4}. 
During the post-pellet phase, the fluctuation level radially drops faster in the high iota FTM configuration (figure \ref{f:fig_6}.right) than in the standard EJM one where the radial reduction is more gradual (figure \ref{f:fig_6}.left).

\begin{figure}[h]
\center
\includegraphics[width=0.49\columnwidth,trim= 0 0 10 0]{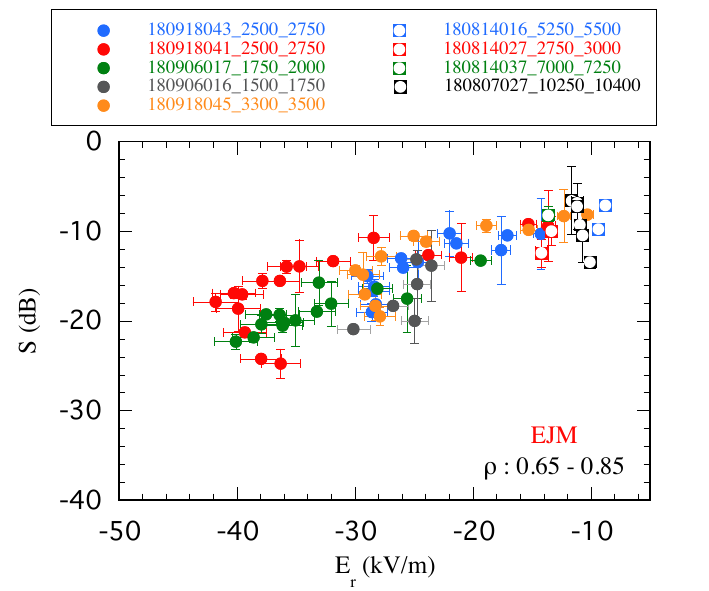}
\includegraphics[width=0.49\columnwidth,trim= 0 0 10 0]{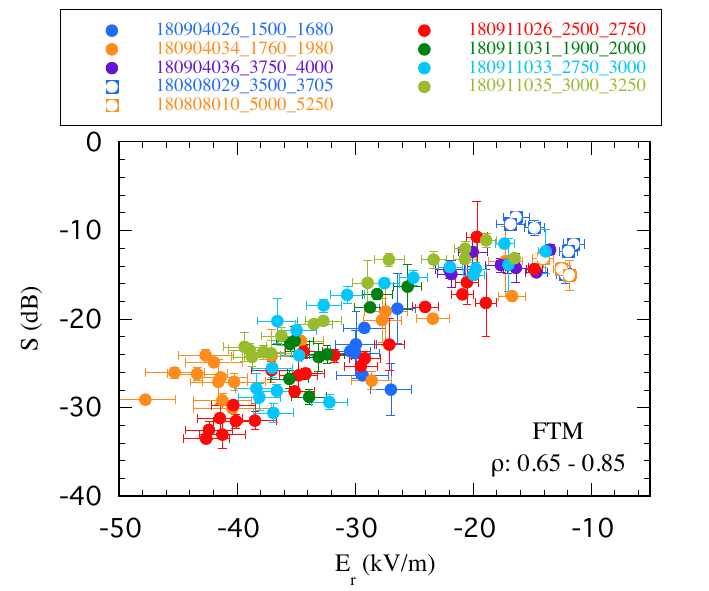}
\caption{Density fluctuation level as a function of local $E_r$ measured within the radial range $\rho=$ 0.65 - 0.85 in several experimental programs in the standard (left) and high iota (right) magnetic configurations. Solid symbols represent data obtained during the post-pellet phase and open symbols in gas fuelled reference programs.}
\label{f:fig_7}
\end{figure}

A similar behaviour is found in all programs studied in this work. This is clearly seen when the local density fluctuation level is represented versus the local $E_r$ intensity as shown in figure \ref{f:fig_7}. 
This representation reflects mainly the radial evolution of the two quantities rather than any direct dependence of the fluctuation level on $E_r$, and clearly shows the differences when the two configurations are compared. A linear trend is visible in both configurations, in the standard EJM (figure \ref{f:fig_7}.left) and in the high iota FTM (figure  \ref{f:fig_7}.right), but with steeper slope in the later. 
The differences found when the two magnetic configurations are compared may be associated directly to configuration properties but also to the differences in the plasma profiles that evolve differently in the post-pellet phase (as can be seen in figure \ref{f:fig_3}.a and \ref{f:fig_3}.c).

\subsubsection{Comparison with gyrokinetic simulations.}

Gyrokinetic simulations using the codes EUTERPE~\cite{Jost:2001,Kornilov:2004,Sanchez:2019,Sanchez:2020} and \texttt{stella}~\cite{Barnes:2019} have been performed aiming to disentangle the impact of $E_r$, plasmas profiles, and magnetic configuration on the density fluctuations.
The simulations are to some extent complementary. \texttt{Stella} simulations consider kinetic electrons but are local (do not consider $E_r$), while EUTERPE simulations are global and with adiabatic electrons. 
The same four scenarios (as those presented in figures \ref{f:fig_3}, \ref{f:fig_4} and \ref{f:fig_6}) are considered: two post-pellet enhanced confinement phase programs with similar density ($\sim 9 \times 10^{19}$ m$^{-2}$) and ECH heating power ($\sim 5.5 - 6.0$ MW) the first in the standard magnetic configuration and the second in the high iota configuration, and the two reference programs with similar plasma density and heating power but fuelled with gas puffing.
As for the neoclassical simulations (shown in figure \ref{f:fig_4}) the electron density and temperature profiles measured by Thomson Scattering diagnostic~\cite{Pasch:2016} and the ion temperature profile measured by XICS diagnostic~\cite{Pablant:2018} (shown in figure \ref{f:fig_3}) are taken as an input.

\begin{figure}[h]
\center
\includegraphics[width=0.32\columnwidth,trim= 10 0 10 0]{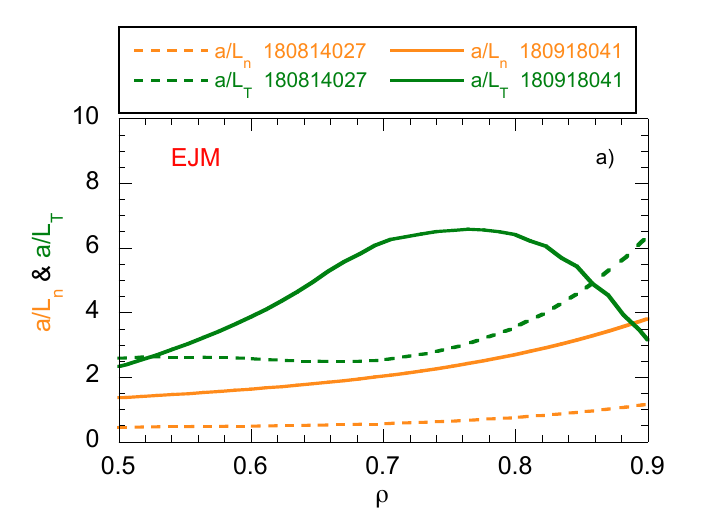}
\includegraphics[width=0.32\columnwidth,trim= 10 0 10 0]{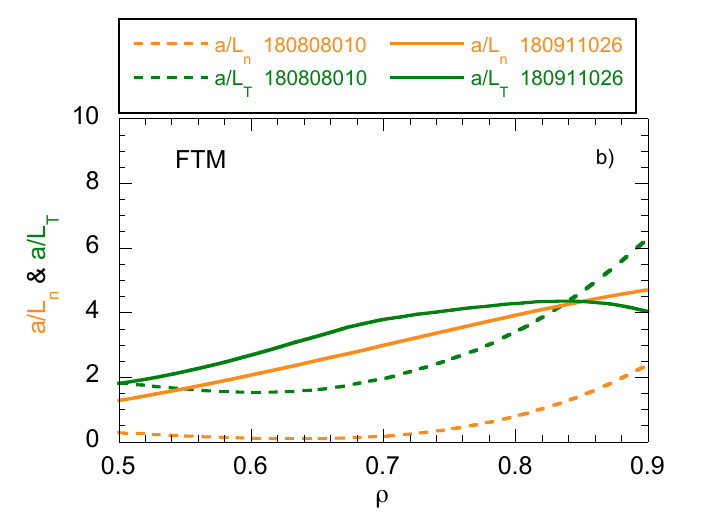}
\includegraphics[width=0.32\columnwidth,trim= 10 0 10 0]{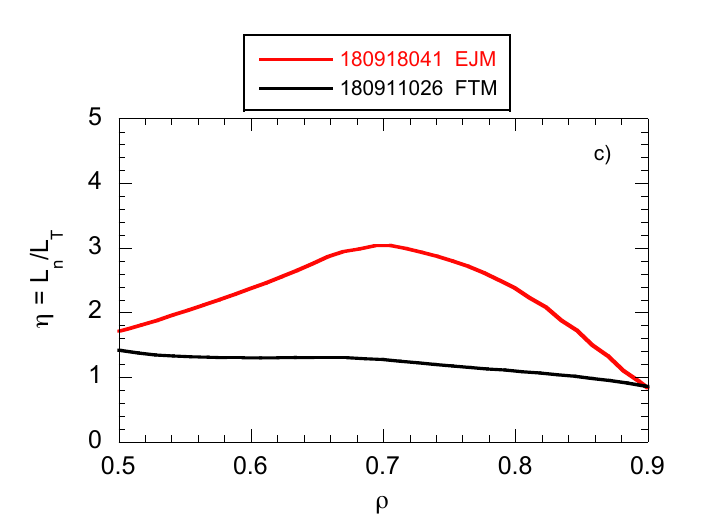}
\caption{Normalised density gradient (in orange) and temperature gradient (in green) for gas fuelled (broken lines) and post-pellet plasmas (solid lines), in the standard EJM (a) and in the high iota FTM (b) configurations. (c): ratio of the temperature gradient and the density gradient for post-pellet plasmas in the standard (in red) and high iota (in black) configurations.}
\label{f:fig_8}
\end{figure}

Local (flux tube) linear gyrokinetic simulations have been performed using the code \texttt{stella}~\cite{Barnes:2019}. These simulations consider collisionless plasmas with kinetic electrons. 
In the simulations, we consider the flux tube centred at $\phi = 0^\circ$, $\theta = 0^\circ$ (where the ITG is expected to be more unstable, see e.g.~\cite{Helander:2012} for a discussion), which lays in the vicinity of the DR measurement region.
The flux tube is located at $\rho \sim 0.7$, a radial position where the modification in the plasma profile gradients reach a maximum ($\rho \sim 0.7$ - $0.75$, see figure \ref{f:fig_8}), and at the same time, the drop in the density fluctuations is highest ($\rho \sim 0.65$ - $0.7$, see figure \ref{f:fig_6}).
The normalised density and temperature gradients displayed in figure \ref{f:fig_8} show that the post-pellet plasmas are characterised by higher density and temperature gradients as compared to the gas fuelled plasmas. In the later, density gradients are rather low in the two configurations resulting in a high $ \eta = L_n / L_T$  $\geq 5$. In the post-pellet plasmas, differences are clearly seen when the two configurations are compared. The evolution of the plasma profiles after the injection of the pellets is different with higher $ \eta$ in the standard EJM than in the high iota FTM configuration (see figure \ref{f:fig_8}.c).

\begin{figure}[h]
\center
\includegraphics[width=0.5\columnwidth,trim= 0 0 0 0]{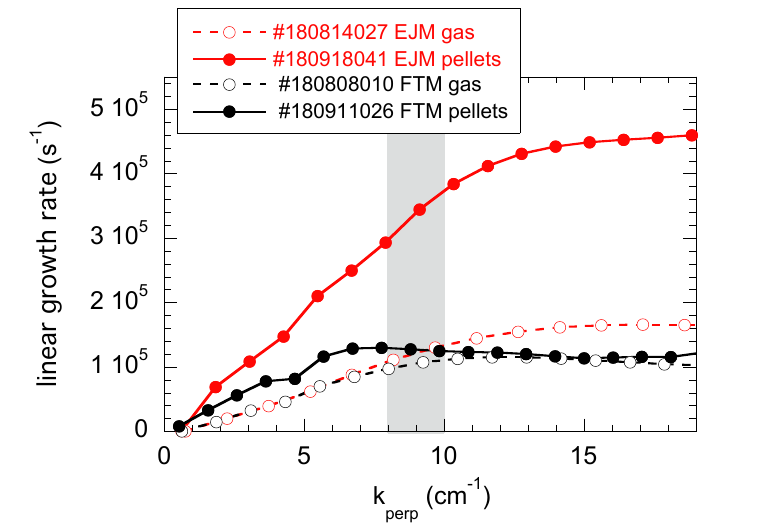}
\caption{Linear growth rates of the instabilities obtained with the code \texttt{stella} for the four experimental programs: post-pellet phase (solid symbols, thick lines) and gas fuelled reference programs (open symbols, broken lines); in the standard configuration (in red) and in the high iota configuration (in black). The grey band indicates the $k_\perp$ range accessible by DR.}
\label{f:fig_9}
\end{figure}

The linear growth rates of the instabilities are shown in figure \ref{f:fig_9}.
For the $k_\perp$ range accessible by DR ($k_\perp \sim 7-10$ cm$^{-1}$), simulations yield similar linear growth rates of the instabilities in the reference gas fuelled plasmas in the two configurations ($\gamma \sim 1 - 1.2 \times10^5$ s$^{-1}$).  In the post-pellet phase plasmas however, a higher growth rate is obtained in the standard EJM configuration ($\gamma \sim  3 - 4 \times 10^5$ s$^{-1}$) while it remains almost unchanged in the high iota FTM configuration. These differences can be partially understood taking into account that the evolution of the plasma profiles after the injection of the pellets is different in the two configurations: steeper temperature gradient in the EJM configuration (turbulence drive) and steeper density gradient in the FTM configuration (turbulence stabilization). Nevertheless, these linear simulations cannot explain the lower density fluctuation level measured during the post-pellet phase plasmas. The simulations also show differences in the parallel mode structure. The localization of the fluctuations along the flux tube peaks in the vicinity of the DR measurement region in the gas fuelled plasmas (characteristic of ITG modes) but moves out of this region in the post-pellet cases (characteristic of TEM modes). This could partially reconcile situations with higher growth rate and ion heat flux and, at the same time, lower local measure of density fluctuation levels. 
Finally, it has to be noted that collisions may also affect the results. As discussed in~\cite{Alcuson:2020}, a reduction in the linear growth rates of the TEM is expected if collisions are included in the simulations. Thus, collisions would affect the growth rates in the post-pellet cases but to less extent those in the gas fuelled plasmas being dominated by ITG modes.

Global linear gyrokinetic simulations have been carried out using EUTERPE to further study the effect of plasma profiles, $E_r$ and magnetic configuration on the instabilities.
These simulations consider collisionless ITG-driven turbulence with adiabatic electrons.
The computational domain is restricted to the radial range $\rho=0.65-0.85$ to make the simulations computationally less expensive.
The results of the simulations show that  the most unstable modes are located at $\rho \sim 0.75-0.85$ and have $k_\perp \sim 9-10$ cm$^{-1}$, except in the post-pellet phase plasmas in the standard magnetic configuration where the modes are located at inner radial positions $\rho \sim 0.65-0.7$ and have higher $k_\perp \sim 12-13$ cm$^{-1}$, slightly out of the range measured by the DR ($k_\perp \sim 7-10$ cm$^{-1}$).
The simulations indicate that the phase velocity of the most unstable modes goes in the ion diamagnetic drift direction (as expected for ion-driven-modes), and oposite to the $E \times B$ drift, with values around 0.2 - 0.3 km/s in the gas fuelled plasmas and up to 0.6 km/s in the post-pellet phase cases. These velocities are significantly smaller than the perpendicular rotation velocities measured in the experiments, supporting the assumption $v_{E\times B} \gg v_{ph}$. According to these results, obtaining $E_r$ from the perpendicular rotation velocity as $E_r = u_\perp B$, represents an error in the determination of the local $E_r$ of about 5$\%$ towards less negative values.

\begin{figure}[h]
\center
\includegraphics[width=0.48\columnwidth,trim= 20 10 20 10]{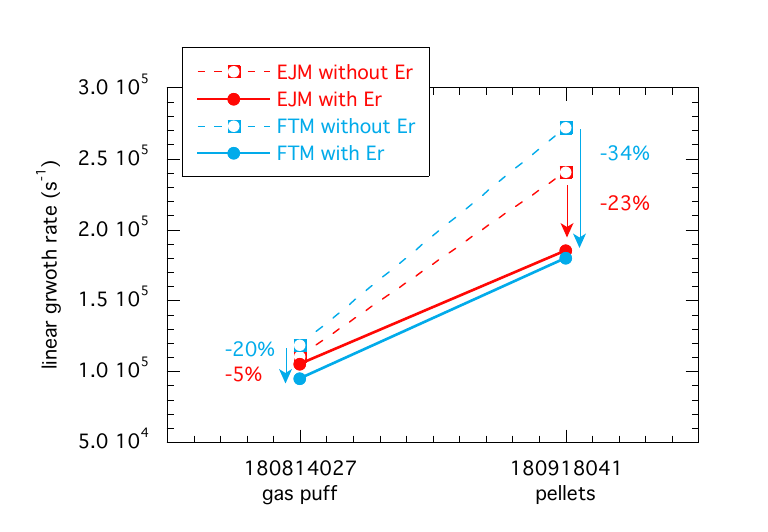}
\includegraphics[width=0.48\columnwidth,trim= 20 10 20 10]{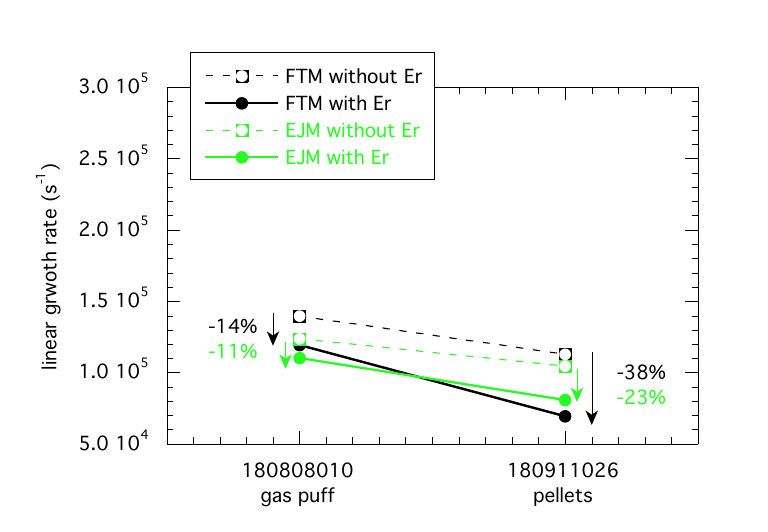}
\caption{Linear growth rates of the most unstable modes obtained in global linear gyrokinetic simulations using EUTERPE for the four experimental programs: two programs in the standard configuration EJM (left, in red) and the other two in the high iota configuration FTM (right, in black), without and with $E_r$ (open and solid symbols, respectively). In blue and green, the growth rates obtained swapping the magnetic configurations. The reduction in the growth rate associated to $E_r$ is indicated for each case.}
\label{f:fig_10}
\end{figure}

The linear growth rates of the most unstable modes for the four experimental programs are shown in table \ref{tab_1} and in figure \ref{f:fig_10}. 
First studies performed ignoring $E_r$ show that in the standard EJM magnetic configuration, the simulations yield a higher linear growth rate of the instabilities in the post-pellet phase plasma as compared to the reference gas fuelled plasma (figure \ref{f:fig_10}.left, open symbols in red), in line with the results of the flux-tube simulations performed with \texttt{stella} (shown in figure \ref{f:fig_9}). 
In the high iota FTM configuration, the linear growth rate is slightly lower in the post-pellet phase as compared to reference plasma (figure \ref{f:fig_10}.right, open symbols in black), and much lower than in the standard EJM configuration.
As already stressed, this can be understood taken into account that the evolution of the plasma profiles after the injection of the pellets is different in the two configurations resulting in lower $ \nabla T_i / \nabla n$ ratio in the high iota FTM configuration as compared to that in the standard EJM configuration (as shown in figure \ref{f:fig_8}.c ).

The comparison of the growth rates obtained with the two codes shows a rather good quantitative agreement in all cases but in the post-pellet case in the standard EJM configuration where the growth rate from \texttt{stella} is higher than that given by EUTERPE. However, as the models used in the simulations are different, a quantitative agreement should not necessarily be expected. 
The flux tube computational domain used in \texttt{stella} simulations cannot be expected to provide exactly the same results as full flux surface or global calculations, in general.
In addition, an important difference between the \texttt{stella} and EUTERPE simulations is related to the model used for electrons. They are fully kinetic in \texttt{stella} cases while are considered adiabatic in EUTERPE’s ones. Thus, TEM instabilities are not captured in EUTERPE simulations, only ITG instabilities.

\definecolor{cyan}{rgb}{0.0, 0.72, 0.92}

\begin{table}[htbp]
\begin{center}
\footnotesize
\begin{tabular}{| c  c | c  c | c  c |}
\hline  
\multirow{2}{*}{Program\#}&\multirow{2}{*}{Fuelling}&\multicolumn{2}{  c | }{\bf{EJM}}&\multicolumn{2}{  c |}{\bf{FTM}} \\
  &  &$\gamma$ w/o $E_r$&$\gamma$ with $E_r$&$\gamma$ w/o $E_r$&$\gamma$ with $E_r$\\
\hline \hline \hline
180814027&gas puff&{\color{red}{1.102 10$^{5}$}}&{\color{red}{1.051 10$^{5}$ (-5$\%$)}}&{\color{cyan}{1.185 10$^{5}$}}&{\color{cyan}{0.946 10$^{5}$} (-20$\%$)}\\ 
\hline
\bf{180918041}&\bf{pellets}&{\color{red}\bf{2.406 10$^{5}$}}&{\color{red}\bf{1.865 10$^{5}$} (-23$\%$)}&{\color{cyan}{\bf{2.720 10$^{5}$}}}&{\color{cyan}{\bf{1.797 10$^{5}$}(-34$\%$)}}\\ 
\hline \hline  \hline
180808010&gas puff&{\color{green}{1.237 10$^{5}$}}&{\color{green}{1.102 10$^{5}$} (-11$\%$)}&1.398 10$^{5}$&1.207 10$^{5}$ (-14$\%$)\\ 
\hline 
\bf{180911026}&\bf{pellets}&{\color{green}{\bf{1.046 10$^{5}$}}}&{\color{green}{\bf{0.808 10$^{5}$}(-23$\%$)}}&\bf{1.129 10$^{5}$}&\bf{0.703 10$^{5}$ (-38$\%$)}\\ 
\hline  
\end{tabular}\\
\caption{\label{tab_1}Linear growth rates (in s$^{-1}$) of the most unstable modes obtained in global linear gyrokinetic simulations using the code EUTERPE for the four experimental programs, without and with $E_r$. In red and black the values obtained considering the actual magnetic configurations and in blue and green those obtained swapping the magnetic configurations. The reduction in the growth rate when $E_r$ is included in the simulations is shown in brackets.}
\end{center}
\end{table}
\normalsize

The effect of $E_r$ on the instabilities has been studied by running the same set of simulations with the $E_r$ profiles obtained in the neoclassical simulations (shown in figure \ref{f:fig_4}). 
In all cases a reduction in the linear growth rate is observed when $E_r$ is included in the simulations. While this reduction is relatively modest in the gas fuelled plasmas (5$\%$ in EJM and 14$\%$ in FTM), it is significant in the post-pellet plasmas (23$\%$ in EJM and 38$\%$ in FTM). This result is to some extent expected being the $E_r$ and its shear more intense in the post-pellet cases.
As a result, in the high iota FTM configuration, the linear growth rate of the instabilities is a factor of $\sim 2$ lower in the post-pellet phase than in the reference case when the effect of the $E_r$ (and $E_r$-shear) is considered. This result is in line with the experimental result showing lower fluctuation level in the post-pellet phase as compared to the reference gas fuelled case. 
In the standard configuration however, despite the effect of $E_r$ (23$\%$ reduction), the linear growth rate in the post-pellet phase is still higher than in the reference gas fuelled plasma, what seems to be in contradiction with the experimental results. It is worth mentioning that the growth rates shown in table \ref{tab_1} and figure \ref{f:fig_10} correspond to the most unstable modes that for the post-pellet phase in the EJM configuration have a $k_\perp$ slightly above the range detected by the DR. Thus, lower $k_\perp$ would have a lower growth rate.

The results obtained in the standard magnetic configuration are in line with previous studies where the impact of $E_r$ on the turbulence was studied using the GENE code under similar conditions~\cite{Xanthopoulos:2020}. It was found that $E_r$ produces a displacement of the fluctuations in the magnetic surface towards regions with lower curvature less favourable for turbulence generation. This effect however was not strong enough to compensate the increase in the fluctuations due to the increase in the ion temperature gradient. 

The comparison of the two configurations in the post-pellet phase conditions yields a lower growth rate of the instabilities in the high iota configuration as compared to the standard one, in agreement with the results of the flux-tube simulations performed with \texttt{stella}. This result is in line with the lower fluctuation level measured during the post-pellet phase in the high iota FTM configuration as compared to that measured in the standard EJM configuration. 
It is worth stressing however, that the lower growth rate in the high iota configuration is not directly linked to the magnetic configuration itself but to the differences in the plasma profiles. 
This conclusion results from the comparison of the growth rates obtained in simulations performed with the actual magnetic configuration and those obtained swapping the magnetic configuration but keeping the plasma profiles (shown in light blue and green in table \ref{tab_1} and in figure \ref{f:fig_10}).  
The simulations without $E_r$ carried out with same profiles and swapping the magnetic configuration show that the high iota FTM configuration is more unstable than EJM for the same profiles. On the other hand, the stabilizing effect of $E_r$ is larger in the FTM configuration than in EJM for the same profiles, resulting in slightly lower growth rates in FTM. This difference however does not explain the lower fluctuation level measured in the FTM configuration, that is rather a consequence of the differences in the plasma profiles of the two pellet fuelled experimental programs ($\#$180918041 and $\#$180911026). Lower growth rates are obtained for the program $\#$180911026 regardless the magnetic configuration considered in the simulations.  

The results of the linear simulations discussed in the present work are in line with the experimental results obtained in the FTM configuration, as the density fluctuation level can be expected to increase with the linear growth rate. 
This is not the case found in the EJM configuration, in which higher linear growth rates but lower density fluctuation levels are found in the post-pellet phase than in the gas fuelled plasma.
However, the fluctuation level is not determined only by the linear growth rate but depends also on the nonlinear saturation of instabilities.
As it has been previously reported, the role of zonal flows on the non-linear saturation of turbulence could yield results not envisaged within a linear framework~\cite{Watanabe:2007}. 
It is well known that zonal flows affect the onset of ITG instabilities~\cite{Dimits:2000a} and it has been shown that magnetic configurations linearly more unstable to ITG modes but with higher zonal flow response may lead to a stronger regulation of the turbulent transport~\cite{Watanabe:2008}. Besides, the zonal flow response has been found to depend on $E_r$~\cite{Sugama:2009,Kleiber:2010,Watanabe:2011}. 
Recently, discrepancies between expectations based on linear simulations and the heat fluxes obtained in non-linear simulations have also been found in W7-X configuration~\cite{Banon_Navarro:2020}. 
Thus, non-linear gyrokinetic analyses could eventually resolve the disaccord found between experimental results and linear simulations. 
The stronger $E_r$ in the post-pellet plasmas, as compared to that in the gas fuelled plasmas, may affect not only the linear growth rates, as found in this work, but also the zonal flow response and eventually the turbulent transport regulation.
Future non-linear simulations are foreseen which will allow for a quantitative analysis of these effects and the comparison with the experimental results.

\section{Summary}  \label{Summary}

Radially resolved density fluctuations and radial electric field measurements have been performed during the post-pellet enhanced confinement phase in W7-X using Doppler reflectometry. Two magnetic configurations have been studied, the standard EJM and the high iota FTM configurations. 
After the injection of a pellet series, a strong negative $E_r$ develops that extends deep into the plasma core, similar in the two configurations. The maximum $E_r$ intensity at the $E_r$-well scales with both density and ECH power level following a similar trend as the plasma energy content. The comparison of experimental $E_r$ profiles and neoclassical predictions obtained using the codes DKES and KNOSOS shows good agreement. The flat $E_r$ profiles measured in reference gas fuelled plasmas, and the $E_r$-well radial position and intensity measured during the post-pellet phase, are rather well reproduced by the neoclassical simulations. 
The density fluctuation level shows a radial drop from the plasma edge toward the plasma core, much more pronounced during the enhanced confinement phase than in reference gas fuelled plasmas.  Besides, the density fluctuation level during the enhanced confinement phase is lower in the high iota magnetic configuration than in the standard one, in the radial range $\rho$: 0.6 - 0.8.   
Gyrokinetic simulations using the codes EUTERPE and \texttt{stella} have been performed aiming to disentangle the impact of $E_r$, plasmas profiles, and magnetic configuration on the density fluctuations. The simulations yield different results for the two post-pellet phase plasmas: lower linear growth rates are obtained in the high iota FTM configuration than in the standard EJM configuration. The differences however are not directly linked to the magnetic configuration itself but rather to the differences in the plasma profiles with lower $ \nabla T_i / \nabla n$  ratio in the high iota FTM configuration as compared to that in the standard EJM configuration. As a consequence, a reduction in the growth rate is found in the post-pellet phase plasma when compared to the gas fuelled reference plasma in the FTM configuration but not in the EJM. In all cases, a stabilization effect linked to $E_r$ is observed. The effect is relatively modest in the gas fuelled plasmas (5$\%$ in EJM and 14$\%$ in FTM), but is significant in the post-pellet plasmas (23$\%$ in EJM and 38$\%$ in FTM). This stabilization however is not strong enough to compensate the increase in the growth rate in the EJM configuration linked to a higher $ \nabla T_i / \nabla n$ ratio. Thus, the simulation results are in line with the experimental results obtained in the FTM configuration but cannot explain those found in the EJM configuration. 
Non-linear simulations are foreseen which will allow for a quantitative comparison with the experimental results.

\section*{Acknowledgements}

\small{The authors acknowledge the entire W7-X team for their support.
This work has been partially funded by the Spanish Ministry of Science and Innovation under contract numbers FIS2017-88892-P and PGC2018-095307-B-100. 
The authors acknowledge the computer resources at Mare Nostrum IV and the technical support provided by the Barcelona Supercomputing Center. 
Part of the simulations were carried out using the Marconi supercomputer at CINECA, from the EUROfusion infrastructure.
This work has been carried out within the framework of the EUROfusion Consortium and has received funding from the Euratom research and training programme 2014-2018 and 2019-2020 under grant agreement No 633053. The views and opinions expressed herein do not necessarily reflect those of the European Commission.}

\section*{References} 

\bibliographystyle{prsty_copia}

\bibliography{/Bibtex_database_copia}

\end{document}